# Characterization of Predictable Quantum Efficient Detector over a wide range of incident optical power and wavelength.

Mikhail Korpusenko[1], Farshid Manoocheri[1], Olli-Pekka Kilpi[2], Aapo Varpula[2], Markku Kainlauri[2], Tapani Vehmas[2], Mika Prunnila[2], and Erkki Ikonen[1,2]

[1]Metrology Research Institute, Aalto University, Espoo, Finland;

[2]VTT Technical Research Centre of Finland Ltd, Espoo, Finland;

Corresponding e-mail: mikhail.korpusenko@aalto.fi

We investigate the Predictable Quantum Efficient Detector (PQED) in the visible and near-infrared wavelength range. The PQED consists of two n-type induced junction photodiodes with $Al_2O_3$ entrance window. Measurements are performed at the wavelengths of 488 nm and 785 nm with incident power levels ranging from 100 µW to 1000 µW. A new way of presenting the normalized photocurrents on a logarithmic scale as a function of bias voltage reveals two distinct negative slope regions and allows direct comparison of charge carrier losses at different wavelengths.  The comparison indicates mechanisms that can be understood on the basis of different penetration depths at different wavelengths (0.77 µm at 488 nm and 10.2 µm at 785 nm). The difference in the penetration depths leads also to larger difference in the charge-carrier losses at low bias voltages than at high voltages due to the voltage dependence of the depletion region.

1. Introduction

Silicon photodiodes are widely used in the wavelength range from 300 to 1000 nm to detect light in various applications. Underpinning spectral responsivity scales based on silicon photodiode working standard detectors are the most straightforward solution for quantitative determination of optical power in these applications. Traceability to SI (International System of Units) has been traditionally established using absolute cryogenic radiometers [1, 2] for calibration of the working standard detectors. Silicon photodiode detectors as primary standards would be attractive because use of cryogenic radiometers requires liquid-helium temperatures and dedicated operation personnel, resulting in high maintenance costs. Predictable Quantum Efficient Detector (PQED) provides such a solution where the spectral responsivity of a silicon detector, operated at room temperature, is determined by fundamental constants, wavelength and a small, predicable correction for reflectance and charge-carrier losses [3-7].

One way to decrease the reflectance losses is to apply a trap detector configuration instead of a single photodiode [8-10]. Charge-carrier losses can be reduced in induced junction photodiodes [8,11,12,13] where the pn junction is produced by the electric field of trapped charge in the photon entrance window layer of the diode. Two induced junction photodiodes of the PQED are aligned in a wedged trap configuration providing a primary standard detector for visible wavelengths [3,4,14]. In addition to calibration of working standard detectors, the PQED can be used in various applications in photometry [15] and in measurements of low optical power when operated at liquid nitrogen temperatures [16,17].

Evaluation of the internal quantum deficiency (IQD) of the PQED has a particular interest. When the recombination losses of charge carriers are small and precisely predicted, the responsivity of the detector can be estimated with low uncertainty. PQEDs made of p-type silicon photodiodes with thick $SiO_2$ coating have been validated relative to absolute cryogenic radiometers and show excellent stability of responsivity over ten years [4, 18]. On the other hand, production of p-type PQEDs requires access to suitable lightly doped p-type silicon wafers and time-consuming coating process. An alternative is to use n-type silicon wafers and $Al_2O_3$ surface layer to produce the induced junction, which offers a simpler photodiode production process [19], well known in photodiode manufacturing industry. Furthermore, a software to predict the IQD of the n-type PQEDs with 100 ppm (parts per million) relative uncertainty was developed and successfully applied, using data from photocurrent vs. bias voltage measurements over a narrow range of incident optical power [19].

A new batch of n-type induced junction photodiodes for PQEDs was produced in this work. Several measurements were carried out to study optical properties of the photodiodes and PQED, such as evaluation of reflectance, IQD, spatial uniformity, and bias-voltage dependent photocurrent (IV curves). Measurements of the n-type PQED were done in the visible and near-infrared spectral region using several optical power levels over a wide range from 100 µW to 1000 µW. The n-type PQED has not been studied before in the near-infrared wavelength range. The PQED can produce a photocurrent with low charge-carrier losses until silicon starts to become transparent at infrared wavelengths approaching 1000 nm. The aim of this study is to broaden our knowledge of PQED operation in the near-infrared region and at power levels approaching the nonlinearity range of the photodiodes. We describe how some features of the IV curves cannot be seen in the linear scale presentation but are only visible in specific logarithmic scale plots.

## 2. Photodiode fabrication and detector assembly

The PQED was constructed of two n-type silicon photodiodes. Schematic cross-section of the photodiodes is presented in figure 1. The photodiodes were fabricated on highly resistive (> 10 kΩ·cm) double side polished 150-mm-diameter and 675-µm-thick n-type silicon substrates. In figure 1, p-diffusion areas represent the diode contacts. The active area of the photodiode is 11 mm x 22 mm. The induced junction of the photodiode is produced by $Al_2O_3$ coating of the silicon substrate. Negative charge in $Al_2O_3$ induces p-type inversion layer [19,20] over the active area of the photodiode. An early description of the induced junction photodiodes can be found in [11].

The fabrication of the photodiodes of this work follows closely the process flow described in [19]. Here, the fabrication started by thermally growing a 400-nm-thick $SiO_2$ layer. The oxide functions both as a screen oxide for the implantation and a field oxide for the device. The p-implantation areas were patterned with photolithography and the oxide was thinned down to 70 nm for the implantation by wet etching with HF. The patterned front side was implanted with boron and the backside with phosphorus. The implanted areas were activated at 1050 °C. Field oxide was removed from the active- and contact-areas by wet etching. A 30-nm-thick $Al_2O_3$ was deposited by atomic layer deposition. The $Al_2O_3$ was removed outside the active area by wet etching. A 300-nm-thick contact aluminium (Al) was sputter deposited and patterned with wet etching. The device was finalized by 20 minutes of annealing at 425 °C in $H_2/N_2$ ambient.

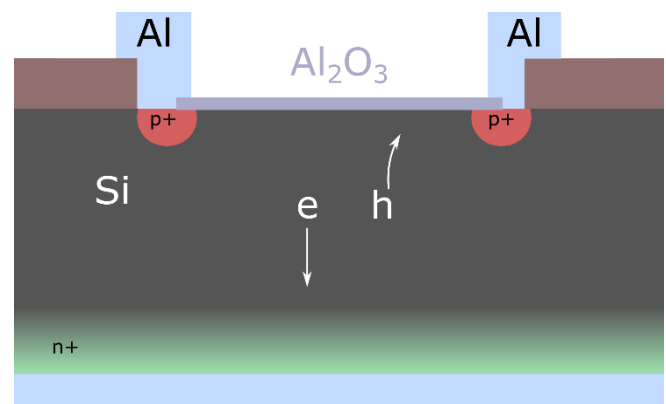

Figure 1. Schematic cross-section of the n-type induced junction photodiode.

The photodiodes were assembled in a light trap configuration as shown in figure 2. The PQED consists of two photodiodes aligned in such a way that seven reflections from the photodiodes take place before the incident beam leaves the detector. The angle between the photodiodes is 15° and the angle of incidence on the first photodiode is 45°. The photodiodes are placed inside a metal cylinder with 10 mm aperture diameter. The PQED is used at room temperature with dry nitrogen flow through the aperture to prevent dust and moisture contamination of the detector. Both photodiodes have own connectors to current measurement electronics which allows detector diagnostics by photocurrent ratio measurement.

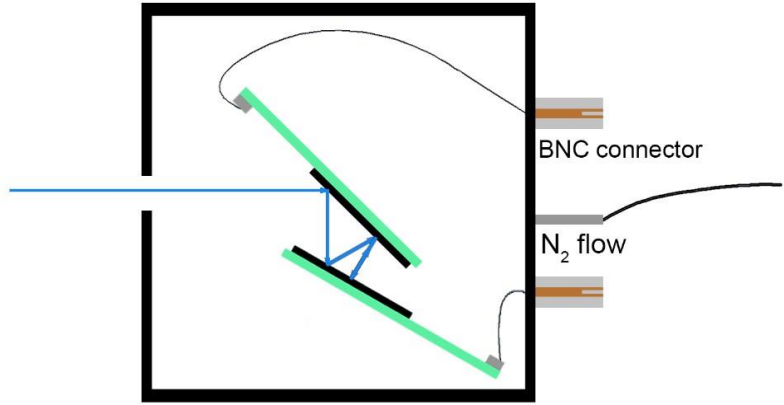

Figure 2. Photodiode assembly and light path in a PQED.

## 3. Characterization measurements and data analysis

### 3.1 Experimental setup

To investigate properties of the n-type PQED we executed following measurements: spatial uniformity scanning of the detector responsivity, detector reflectance measurements, responsivity measurements against p-type PQED and measurement of photocurrent dependence on bias voltage at various incident power levels. The measurement setup (figure 3) includes two laser sources: an argon-ion laser at 488.12 nm wavelength with a Gaussian-like beam diameter of 1.3 mm ($1/e^2$) and a single longitudinal mode semiconductor laser at 784.83 nm wavelength with a diameter of 2.6 mm of the vertically polarized beam. Both lasers were simultaneously used only when aligning the 785 nm laser beam for reflectance measurements. An optical power stabilizer provided a stable p-polarized laser beam, and a wedge mirror with monitor detector was used for laser drift correction. Tested detectors were placed on a moving XY-stage to execute automated measurements. The PQEDs were reverse biased and the sum of the photocurrents from two photodiodes was recorded.

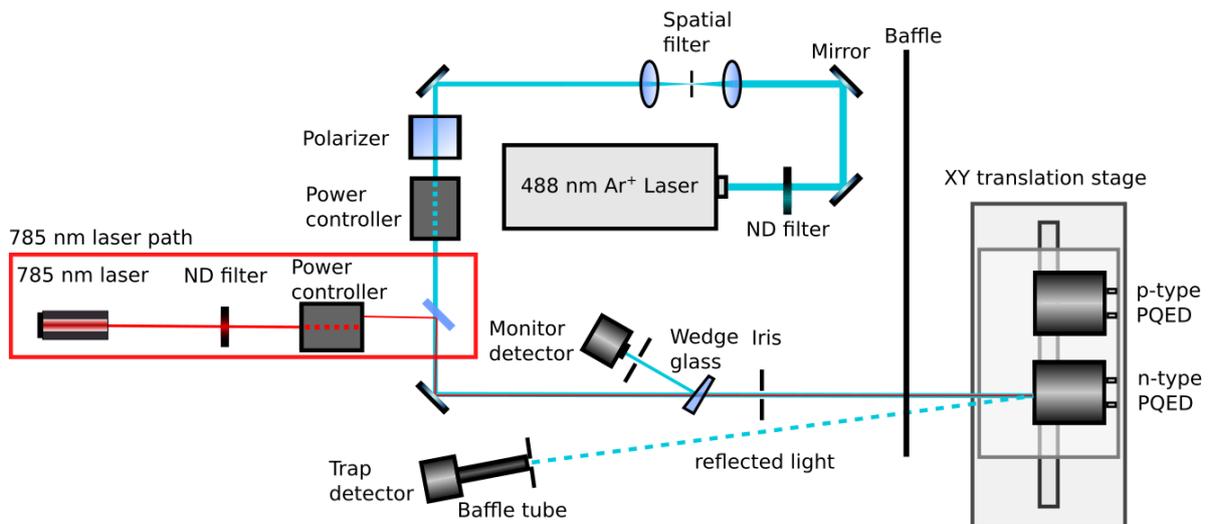

Figure 3. Block diagram of the measurement setup.

## 3.2 Spatial uniformity scanning

To evaluate the spatial uniformity of responsivity of the PQED, we made a scanning measurement of the detector. Here the 488 nm laser beam was used because of smaller diameter. The PQED was placed on the XY translational stage which moved at 0.5-mm steps in vertical and horizontal directions. Figure 4 shows that the uniformity of responsivity is about 60 ppm in the central area with the size of 2 mm x 1 mm. There are two small areas of reduced responsivity on the left side of the detector. They may be caused by dust particles or defects in the detector structure. Spatial uniformity of another n-type PQED within 30 ppm in the area of 4 mm in diameter has been measured in [19].

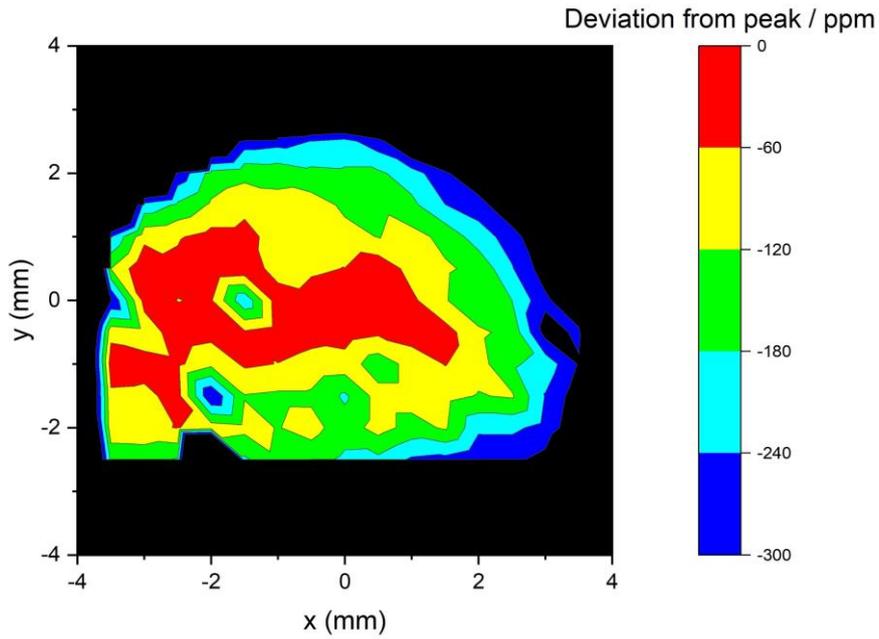

Figure 4. Spatial uniformity of PQED responsivity.

### 3.3 Reflectance and responsivity

A p-type PQED was used as the reference detector in responsivity measurements of the n-type PQED of this study. The PQEDs were placed on a moving stage and the setup automatically measured photocurrents $I_p$ and $I_n$ from the p- and n-type detectors, respectively. Each detector was connected to a separate current-to-voltage converter (CVC) with a reverse bias voltage of 5 V. The incident optical power was about 100 μW. During data processing, dark currents and offsets of multimeters and CVCs were corrected. The ratio $I_n/I_p$ of the corrected photocurrent values is equal to the responsivity ratio of the PQEDs, when the same optical power is measured by both detectors.

The spectral responsivity of the PQED is given by

$$R(\lambda) = R_0(\lambda) [1 - \rho(\lambda)] [1 - \delta(\lambda)] \tag{1}$$

where $R_0(\lambda) = e\lambda/hc = (\lambda/\mu m)/(1.23984 \text{ W/A})$ is the responsivity of an ideal quantum detector expressed by the vacuum wavelength $\lambda$ of the incident radiation and fundamental constants $e$, $h$, $c$. Parameters $\rho(\lambda)$ and $\delta(\lambda)$ describe the spectral reflectance of the PQED and charge-carrier losses (IQD) of the photodiodes, respectively. To determine the difference of internal charge-carrier losses $\delta_n(\lambda) - \delta_p(\lambda)$ of the n-type and p-type photodiodes, the detector reflectance values need to be taken into account according to equation (1).

The reflectance of the detectors was measured with the use of a calibrated Hamamatsu silicon trap detector with a baffle tube as shown in figure 3. For both laser wavelengths the optical power in reflectance measurements was around 1000 µW. The reflected beam of the 488 nm laser can be observed with bare eye if the laser power is large enough. Since the reflection of the 785 nm beam is too weak to be detected by bare eye or luminescence cards, an extra step was applied. We used the 488 nm laser as an auxiliary beam and ensured that the paths of these two beams, passing the PQEDs, totally coincide over a long distance. After reflection from the PQED, the reflectance at 785 nm could be measured when blocking the 488 nm laser beam. In these measurements 10 V reverse bias voltage was applied to the detector because of the large incident power.

| Laser wave-length (nm) | Photocurrent ratio of PQEDs, $I_n/I_p$ | Reflectance of p-type PQED, $\rho_p$ (ppm) | Reflectance of n-type PQED, $\rho_n$ (ppm) | IQD correction to the beam diameter of 1.3 mm (ppm) | IQD difference $\delta_n - \delta_p$ (ppm) |
|---|---|---|---|---|---|
| 488.12 | 0.999722 (±20 ppm) | 25 ± 1 | 117 ± 1 | 0 | 185 ± 20 |
| 784.83 | 0.999664 (±20 ppm) | 45 ± 8 | 71 ± 1 | -46 ± 10 | 264 ± 24 |

Table 1. Photocurrent ratio, reflectance, beam diameter correction, and internal quantum deficiency (IQD) difference of n-type and p-type PQEDs. Standard uncertainty of the photocurrent ratio measurements is 20 ppm. Other reflectance values are measured, but the reflectance of p-type PQED at 785 nm is obtained from a validated calculation [3] with a standard uncertainty of 8 ppm.

Table 1 shows the measured photocurrent ratios $I_n/I_p$, reflectances for the two wavelengths, and a correction due to the beam size difference caused by the nonuniformity of spatial responsivity (figure 4). Reflection loss values agree well with previous measured and calculated results [3,6,19]. For small reflectance and IQD values, equation (1) can be used to approximate the photocurrent ratio as

$$I_n/I_p \approx 1 + \rho_p(\lambda) - \rho_n(\lambda) + \delta_p(\lambda) - \delta_n(\lambda) . \qquad (2)$$

The rightmost column in Table 1 gives the resulting differences in the IQD values. It is clearly seen that the n-type PQED has larger charge-carrier losses than the p-type PQED. Furthermore, these losses are larger by (79 ± 31) ppm at 785 nm than at 488 nm wavelength. Estimates of absolute IQD of the n-type PQED are (188 ± 73) ppm at 488 nm and (267 ± 65) ppm at 785 nm, where the standard uncertainty is dominated by the uncertainty of the predicted responsivity of the p-type PQED [3,5]. The values include a 6 ppm correction for the estimated absorption loss in the $Al_2O_3$ layer [19].

### 3.4 Photocurrent dependence on bias voltage

To study recombination losses in the photodiodes, measurements of photocurrent as a function of bias voltage were carried out. We used reverse bias voltages between 0 and 18 V to record the change of the detector photocurrent when applying a constant optical power between 100 µW and 1000 µW to the PQED. As the bias voltage source, we used a Keithley calibrator, which can provide a very stable voltage signal in the range of ±20 V.

In the preliminary data analysis, the highest measured photocurrent was used as a reference value. Normalized photocurrent data are presented as $q = I(V)/I_{max}$ where $I(V)$ is the photocurrent with the applied bias voltage $V$ and $I_{max}$ is the largest of the recorded current values at a certain optical power level. However, the linear data plots of figures 5(a) and 5(b) do not provide a sufficient view to the overall behavior of the photocurrent values at all bias voltages.

A better view is obtained on the logarithmic scale which the selected normalization allows to use in the form $Q = \log(1 - q)$. The data of figures 5(c) and 5(d) on the logarithmic scale show two distinct negative slope regions as a function of the bias voltage. These characteristics are not easily visible on the linear scale. For the curves corresponding to 1000 µW power level, the slope changes at the bias voltage of 2.5 V for 488 nm and at 3 V for 785 nm. At 180 or 190 µW power, the corner point is at about 0.5 V at both wavelengths.

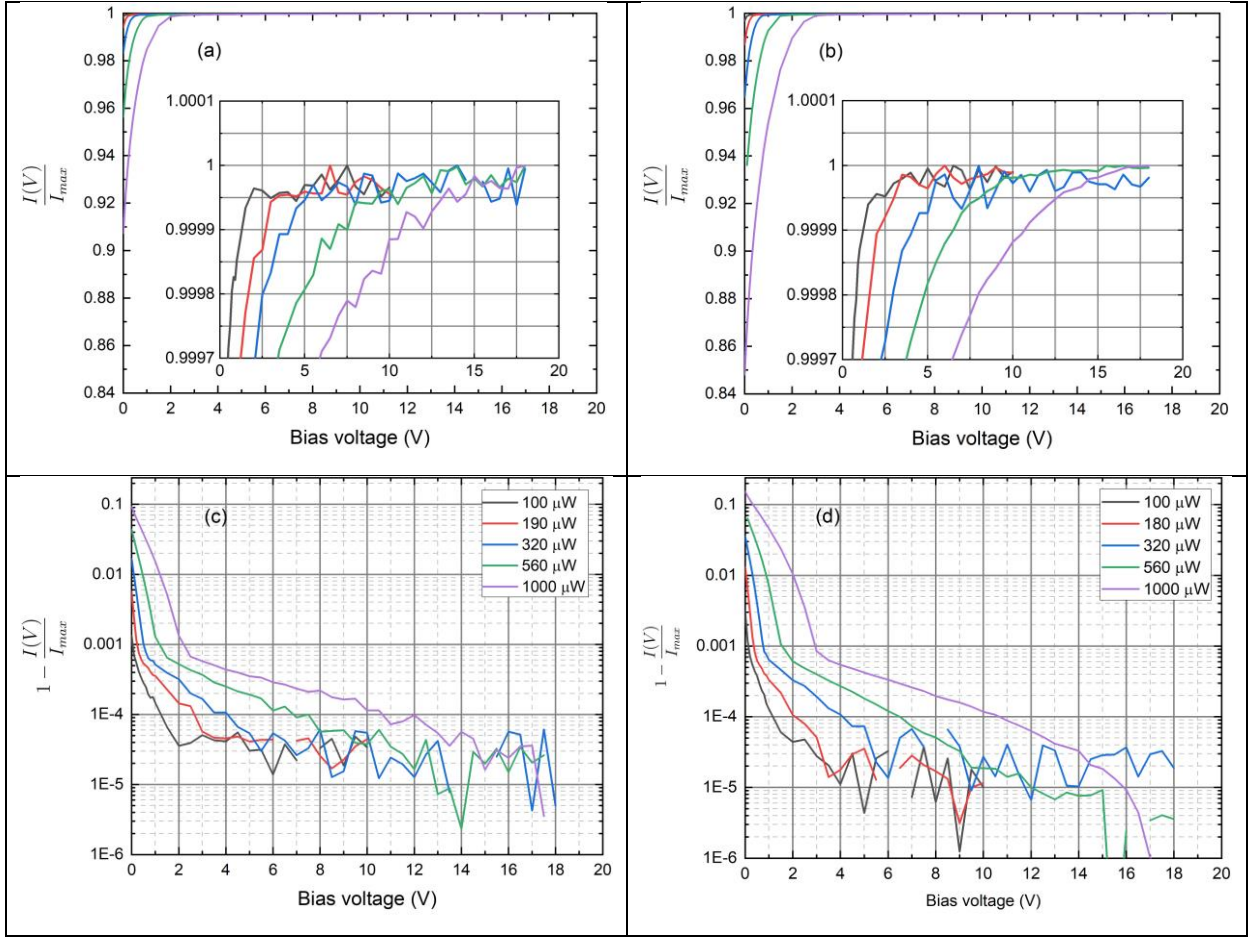

Figure 5. Photocurrent dependence on applied reverse bias voltage for 488 nm laser beam in linear (a) and logarithmic scale (c) and correspondingly for the 785 nm laser beam [(b) and (d)]. The data points corresponding to $I(V) = I_{max}$ cannot be presented on the logarithmic scale.

Selection of $I_{max}$ to normalize the photocurrent values is practical but rather arbitrary. The question of a proper normalizing photocurrent can be addressed by multiplying both sides of equation (1) with the incident optical power $P$. Noting that $R(\lambda)P = I(V)$ and solving for IQD gives

$$\delta(\lambda) = 1 - I(V)/I_{sat}(\lambda), \qquad (3)$$

where $I_{sat}(\lambda) = [1 - \rho(\lambda)]\, R_0(\lambda)P$ is the saturation photocurrent of an otherwise ideal photodetector, except that $\rho(\lambda) > 0$. Equation (3) indicates that using $I_{sat}(\lambda)$ as the normalizing photocurrent instead of $I_{max}$ makes the vertical scale of the IV curves to correspond to IQD. In figures 5(a) and 5(b), the location of $I_{sat}(\lambda)/I_{max}$ is approximated to be 188 ppm and 267 ppm above the average level of saturated $I(V)/I_{max}$ values, respectively, corresponding to the absolute IQD values given at the end of section 3.3. The average saturated levels can be estimated to be 0.99998 at both 488 nm and 785 nm.

Using the normalization of equation (3) on the logarithmic scale, the photocurrent dependence on bias voltage is reproduced in figure 6 in such a way that the curves at different wavelengths can be easily compared with each other. It is seen that on the logarithmic scale there is an approximately constant difference between the curves at different wavelengths, especially at voltages above the corner point. That conclusion would change if $I_{max}$ would be used as the normalizing photocurrent because the curves would overlap at high bias voltages.

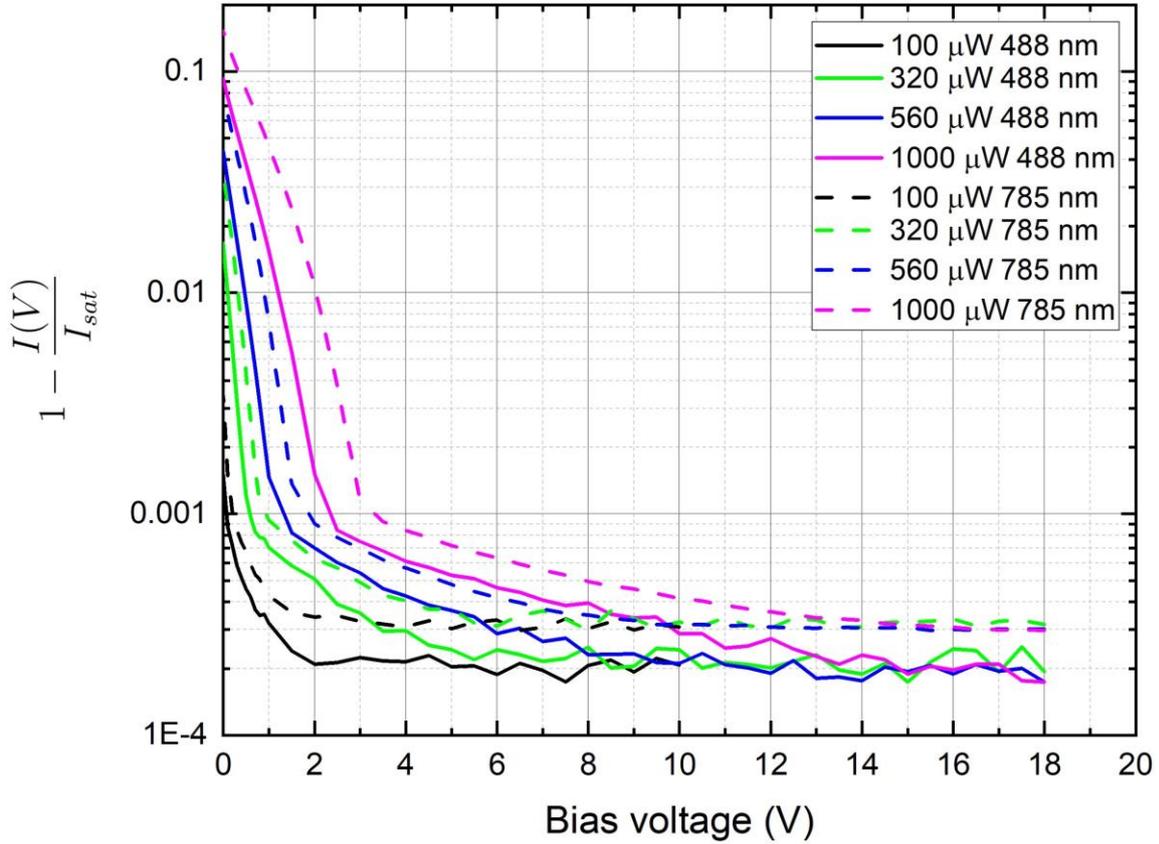

Figure 6. Comparison of photocurrent dependence on applied reverse bias voltage for 488 and 785 nm laser light at different levels of incident power. Note that use of the saturated photocurrent of an ideal quantum detector ($I_{sat}$) for normalization, instead of $I_{max}$ of figure 5, allows correct description of the differences of charge carrier losses at high bias voltages.

## 4 Discussion

The measured IQD values at 488 nm are similar as reported in [19] for n-type PQED, but a good spatial uniformity is obtained over a somewhat smaller area. Furthermore, this work indicates that the IQD at infrared wavelengths appears to be considerably higher in n-type PQED than in good quality p-type PQED.

The purpose of bias voltage dependence experiments was to detect the saturation point of the photocurrent at the measured wavelength depending on the power level. In addition, comparison of $I(V)/I_{sat}(\lambda)$ ratios at a fixed bias voltage can evaluate linearity of the detector depending on the incident optical power. For example, it can be seen from figure 6 that in the conditions of the reflectance measurements (1000 µW power, 10 V bias voltage), the additional charge-carrier losses relative to the saturated IQD are smaller than 0.02 %. Such deviation causes nonlinearity in the responsivity, but the low value of such nonlinearity does not affect here the reliability of the reflectance measurements of the n-type PQED.

Crucial parameters in determining the logarithmic IV curves are laser stability and noise level. In most of these measurements the noise level did not exceed 60 ppm. Initially the saturation photocurrent of an ideal quantum detector to be used for normalization of IV curves is unknown. In this work, comparison with p-type PQED allowed to assign numerical values to $I_{sat}(\lambda)$ of n-type PQED. It is expected that properly normalized logarithmic IV curves are useful in fitting three-dimensional charge-carrier recombination models to the experimental data, with the final goal of low-uncertainty determination of the IQD of PQED photodiodes [21].

The increase of IQD with increasing wavelength for n-type PQED was measured for the first time in this work and needs to be discussed. The penetration depths are 0.77 µm at 488 nm wavelength and 10.2 µm at 785 nm [22]. With exponential decay, only 2 ppm of incident power remains at 488 nm wavelength at the depth of 10 µm inside the photodiode. Thus all 488 nm light is absorbed within the depletion region, the width of which is calculated to be approximately 60 µm at 0 V bias voltage, 100 µm at 1 V bias and 350 µm at 20 V bias. For the wavelength of 785 nm, the situation is different at low bias voltages, because 3500 ppm and 80 ppm of incident power remains at the depth of the depletion region width at 0 V and 1 V bias voltages, respectively. Those values provide a qualitative explanation of figure 6 which indicates that the differences between charge-carrier losses at 785 nm wavelength and at 488 nm are larger at low bias voltages than at high voltages.

## Acknowledgements
Hannu Ronkainen is acknowledged for useful discussions. We thank financial support from Business Finland co-innovation project RaPtor (Project no. 6030/31/2018). This work is partly funded by the Academy of Finland Flagship Programme, Photonics Research and Innovation (PREIN), decision number: 320167.